\documentclass[10pt,preprint]{sigplanconf}
\usepackage{multirow}
\usepackage{amsmath}
\usepackage{amssymb}
\usepackage{empheq}
\usepackage[vlined,linesnumbered,ruled]{algorithm2e}
\usepackage{fancyvrb}
\usepackage{color}
\usepackage{booktabs}
\usepackage{hyperref}
\usepackage[normalem]{ulem}
\usepackage{verbatim}
\usepackage{listings}
\usepackage[labelformat=simple]{subcaption}

\usepackage{cancel}
\usepackage{appendix}
\usepackage{lmodern}

\usepackage{setspace}
\DeclareMathAlphabet\mathbfcal{OMS}{cmsy}{b}{n}

\definecolor{dmlgreen}    {RGB}{51,  160,  44}
\definecolor{dmlblue}     {RGB}{31,  120, 180}
\definecolor{dmlred}      {RGB}{202,   0,  32}
\newsavebox{\verbsavebox}
\newcommand{\ttt}{\texttt}

\newcommand{\REM}[1]{}

\hypersetup{draft}

\begin{document}
\title{Building Application-Specific Overlays on FPGAs with High-Level Customizable IPs}
\authorinfo{Hongbo Rong}{Parallel Computing Lab (PCL), Intel}{hongbo.rong@intel.com}

\maketitle

\begin{abstract}

Overlays are virtual, re-configurable architectures that overlay on top of physical FPGA fabrics~\cite{fpgaOverlays16}. An overlay that is specialized for an application, or a class of applications, offers both fast reconfiguration and minimized performance penalty. Such an overlay is usually implemented by hardware designers in hardware ``assembly'' languages at register-transfer level (RTL).
 
This short article proposes an idea for a {\it software programmer}, instead of hardware designers, to quickly implement an application-specific overlay using high-level customizable IPs. These IPs are expressed succinctly by a specification language, whose abstraction level is much higher than RTL but can nonetheless expresses many performance-critical loop and data optimizations on FPGAs, and thus would offer competitively high performance at a much lower cost of maintenance and much easier customizations. 

We propose new language features to easily put the IPs together into an overlay. A compiler automatically implements the specified optimizations to generate an efficient overlay, exposes a multi-tasking programming interface for the overlay, and inserts a runtime scheduler for scheduling tasks to run on the IPs of the overlay, respecting the dependences between the tasks. While an application written in any language can take advantage of the overlay through the programming interface, we show a particular usage scenario, where the application itself is also succinctly specified in the same language.

We describe the new language features for expressing overlays, and illustrate the features with an LU decomposer and a convolutional neural network. A system is under construction to implement the language features and workloads.
\end{abstract}

\section{Introduction}
\label{sec:intro}

An FPGA has massive amount of logical elements that are distributed, locally connected and running in parallel, interleaved with memory blocks and often with hardened DSP blocks. The logical elements, interconnects, memory and DSP blocks can be synthesized to match a dataflow compute for the best performance and power efficiency. However, the synthesis time tends to be very long: even a small design may take tens of minutes, and a larger design can easily take hours or even days.   

Overlays have been proposed to cut down the synthesis time. Overlays are virtual, re-configurable architectures that overlay on top of physical FPGA fabrics~\cite{fpgaOverlays16}. An overlay usually has (much) coarser granularity, and thus (much) smaller amount, of resources that can be re-configured. Therefore, the resources of an overlay can be synthesized for a dataflow compute at a radically faster speed than the traditional hardware synthsis~\cite{fpgaOverlays16,fastHLS14,synthesisFreeJIT11}. 

\REM{An overlay offers software programmers a software-like programming experience and leaves the implementation of the overlay itself to hardware designers, and thus effectively separates the concerns of software and hardware developers at some cost of performance.
}

An overlay offers software programmers a software-like programming experience: An overlay is built with hardware IPs on top of an FPGA; the hardware IPs have a higher-abstraction level (e.g. matrix or vector level), and thus programmers can program the overlay at that higher abstraction level instead, reaching higher productivity at a reasonable performance cost. 

However, there are remaining problems: 
\begin{itemize}
    \item An overlay itself is usually still implemented at RTL, and by hardware experts, with a high development cost.
\item Overlays are often available only for hot domains or applications (e.g. deep learning these days~\cite{brainWave18,vta19,DLA17}). Existing overlays might not necessarily well match new algorithms, applications or domains. 
\end{itemize}

This short article proposes an idea to enable a {\it software programmer}, instead of hardware experts, to quickly build an application-specific overlay on an FPGA using high-level customizable IPs. These IPs are succinctly {\it specified}: the dataflow of an IP is expressed in a functional notation, followed by a description how to efficiently map the dataflow onto the spatial FPGA architecture with many loop and data optimizations, e.g. how to map the dataflow onto a systolic array that well matches the underlying FPGA architecture and thus is critical for performance. 

The IPs are only specified, while the detailed implementation of the specified optimizations is left to a compiler. The specification language and compiler used is T2S (Temporal To Spatial)~\cite{Rong:2018:T2S:arxiv}. Our previous work on T2S has proved that a smart compiler can generate efficient IPs with a fraction of development time but with competitive performance, compared with the same IPs that are optimized in the same set of optimizations but the optimizations are implemented manually by experts in high-level synthesis (HLS) languages~\cite{T2SFCCM19,Susy}~\footnote{We believe that {\it if the compiler is engineered right, the performance of an IP will be mainly determined by the set of optimizations used for the IP, not by whether the optimizations are automatically implemented or manually implemented}. This belief has been supported by our current prototypes.  
Our current prototypes generates HLS code only, and thus we compare only with expert-written HLS code. However, there is no restriction for our approach to generate RTL code, which is purely an engineering effort. 
When generating RTL code, we believe the same phenomenon will repeat: IPs specified in our language and implemented in detail by the automatic compiler should exhibit competitive performance vs. expert-written RTL code with the same set of optimizations. We will verify this belief in future.}.

Since the IPs are written at an abstraction level much higher than RTL, the IPs require much lower maintenance cost, are much easier to customize by software programmers, and on the hand, with the right set of optimizations, can offer competitively high performance. 

The compiler will automatically expose a multi-tasking programming interface for an overlay, and insert a runtime scheduler for scheduling tasks to run on the overlay, respecting the dependences between the tasks. While an application written in any language can take advantage of the overlays through the programming interfaces, we show a particular usage scenario, where the application itself is also succinctly specified in the same language.

This approach is generally applicable to many applications that have many tasks and the tasks need share limited FPGA resources. We will illustrate the approach with a VGG convolutional neural network and a blocked LU decomposer. We will define an overlay for each of them; each overlay contains a few IPs on an FPGA. For the neural network, we map and schedule all the layers to an overlay. For the blocked LU decomposer, we dynamically generate tasks, and schedule them to the other overlay.

This article focuses on describing the idea. We are building a prototype to implement the proposed idea, leveraging our current systems~\cite{T2SFCCM19,Susy}. We will report the progress in future publications.

\section{Overall Flow}

\begin{figure*}[!htb]
    \centering
    \includegraphics[width=0.9\textwidth]{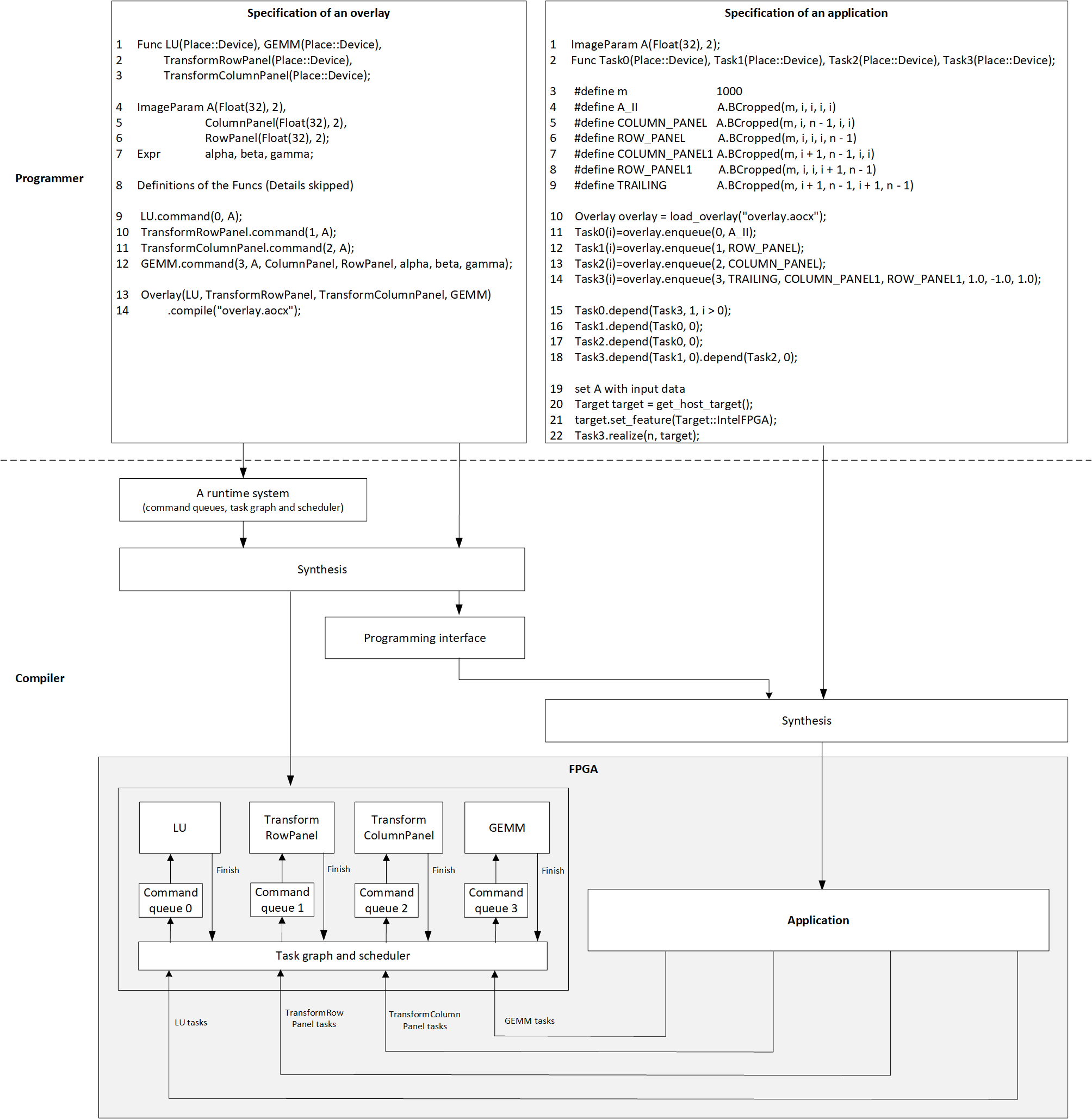}
    \caption{The overall flow}
    \label{fig:T2S_Overlay-overall-flow}
\end{figure*}

Fig.~\ref{fig:T2S_Overlay-overall-flow} shows the overall flow. A programmer specifies a definition of an overlay. Directed by the specification, a compiler automatically links the overlay definition with a pre-written runtime system and synthesize them into a bitstream for an FPGA, and generates a programming interface for the overlay. The runtime system includes command queues, a task graph and a scheduler.

The overlay generated on the FPGA can be invoked to run by an application written in any language by calling the programming interface. A particular interesting scenario is that the application is also written in the same specification language. In Fig.~\ref{fig:T2S_Overlay-overall-flow}, we show that a programmer specifies an application to run on the overlay. The compiler synthesizes the application with the programming interface into another bitstream.

Then the compiler offloads both the overlay and the application to an FPGA. When the programmer invokes the application to run, the application automatically generates tasks for the runtime system to schedule to run on the overlay. The example application shown in the figure is an LU decomposer, which has many tasks of 4 kinds generated during the execution, dispatched by the runtime to run on the 4 corresponding hardware IPs in the overlay. We will describe this example in more detail below.

\section{Examples}

In this section, we illustrate our idea with an LU decomposer and VGG convolutional neural network. Instead of using formal definitions, we will intuitively and effectively explain our language features through these examples.

\subsection{Example 1: Blocked LU decomposition}
\label{sec:lu}

For a matrix $
    A = \begin{pmatrix}
    A_{00} & A_{01} \\
    A_{10} & A_{11} \\
    \end{pmatrix}
$, we would like to decompose it into $A=LU=\begin{pmatrix}
    L_{00} & 0 \\
    L_{10} & L_{11} \\
\end{pmatrix} \begin{pmatrix}
    U_{00} & U_{01} \\
    0 & U_{11} \\
\end{pmatrix}$. 
Therefore, it is easy to see that
\begin{align}
    A_{00} &=& L_{00}U_{00}\\
    A_{01} &=& L_{00}U_{01}\\
    A_{10} &=& L_{10}U_{00}\\
    A_{11} &=& L_{10}U_{01}+L_{11}U_{11}
\end{align}
Therefore,
\begin{align}
    A_{00} &=& L_{00}U_{00}\\
    U_{01} &=& L_{00}^{-1}A_{01}\\
    L_{10} &=& A_{10}U_{00}^{-1}\\
    L_{11}U_{11} &=& A_{11} - L_{10}U_{01}
\end{align}

We can generalize  this example. Suppose the original square matrix $A$ is divided into $n*n$ sqaure blocks, each block having $m*m$ elements. The algorithm of blocked LU is shown in Algorithm~\ref{alg:blu}.

\begin{algorithm}[!htb]
\begin{lstlisting}[
    language=C,basicstyle=\tt,numbers=left,stepnumber=1,showstringspaces=false,tabsize=1,breaklines=true,escapechar=`]
    for ($i=0; i < n; i$++)
      Task 0: decompose $A_{ii}=L_{ii}U_{ii}$ `\label{LU-step1}`
      Task 1: calculate $U_{i, (i+1):n} = L_{ii}^{-1}A_{i,(i+1):n}$ `\label{LU-step2}`
      Task 2: calculate $L_{(i+1):n, i} = A_{(i+1):n, i}U_{ii}^{-1}$
      Task 3: calculate $A_{(i+1):n, (i+1):n}$ -= $L_{(i+1):n,i}U_{i, (i+1):n}$
\end{lstlisting}
\caption{The blocked LU decomposition algorithm.}
\label{alg:blu}
\end{algorithm}

We vision that a T2S specification can be written as shown in Fig.~\ref{fig:T2S_Overlay-overall-flow}. There are 4 hardware IPs:
\begin{itemize}
    \item \ttt{LU}, which accepts a square block $A$ with the size of $m*m$, and decomposes it into matrix $L$ and $U$, and store them at the same space of $A$. Note the diagonal of $L$ contains only 1's, and thus not stored. \\
    \item \ttt{TransformRowPanel}, which accepts a row panel with a number of blocks, each block with the size of $m*m$, and uses the first block (corresponding to $L_{ii}$) to transform the other blocks, i.e. $L_{ii}^{-1}A_{i,(i+1):n}$. \\
    \item \ttt{TransformColumnPanel}, which accepts a column panel with a number of blocks, each block with the size of $m*m$, and uses the first block (corresponding to $U_{ii}$) to transform the other blocks, i.e. $A_{(i+1):n, i}U_{ii}^{-1}$. \\
    \item \ttt{GEMM}, which accepts a matrix $C, A, B$ and co-efficient $\alpha, \beta, \gamma$, and computes $C = \alpha C + \beta A * \gamma B$.
\end{itemize}
All the 4 IPs do in-place update: they write their outputs into the same space of their inputs. 

In Fig.~\ref{fig:T2S_Overlay-overall-flow}, the specifications use several features new to the T2S language:

\begin{itemize}
    \item The \ttt{Overlay} type is a container for the IPs and runtime system.
    \item \ttt{F.command(queueNo, parameters)} specifies a programming interface for Func \ttt{F}: the command queue and the parameters.
    \item \ttt{O.enequeue(queueNo, parameters)} is to enqueue a command to the given command queue of the overlay \ttt{O} with the given parameters. 
    \item \ttt{T1.depend(T2, d, [condition])} says that under an optional \ttt{condition}, task \ttt{T1} in the current iteration depends on task \ttt{T2} in \ttt{d} iterations before.
   \item \ttt{A.BCropped(m, startRow, endRow, startCol, endCol)} means to crop, in blocks of $m*m$, from a buffer \ttt{A}, from the given start to end row (included), and from the given start to end column (included). The cropping is in-place, and thus the cropped buffer shares the space with the original buffer.
\end{itemize}

We can explain Fig.~\ref{fig:T2S_Overlay-overall-flow} in more detail. A software programmer writes two specifications, one for the overaly, and the other for the application (i.e. LU decomposer). 

In the specification of the overlay, Line 1-3 declare the 4 IPs on an (FPGA) device. Line 4-7 declare the inputs of the IPs. Line 8 defines the IPs. We assume that the IPs have already been specified with necessary optimizations in the T2S language by experts, and are provided to the programmer as a library of building blocks. Therefore, we skip the details of the definitions of the IPs here. Line 9-12 define a programming interface for each IP. Each is driven by a command queue, which is automatically provided by a runtime system. Finally, Line 13-14 put the IPs into an overlay, and compile the overlay to a named bitstream.

In the specification of the application, Line 1-2 declare the matrix to be decomposed, and 4 kinds of tasks corresponding to the 4 IPs. Line 3-9 defines some macros that are only for the convenience of usage next. Line 10 offloads the overlay's bitstream to an FPGA, if
not yet, and returns a handle. Line 11-14 generate 4 tasks and enqueue them into the  command queues of the corresponding IPs. Note that there is an implicit loop \ttt{i} around the tasks. In this way, Algorithm~\ref{alg:blu} is expressed. Linie 15-18 specify the dependences between the tasks. Line 19-22 set up the input matrix, compile the application into a bitstream, and run it on the FPGA.

The two specifications are compiled to run on the same FPGA. The compiler will automatically generate for the overlay specification a programming interface, which is used for compiling the application specification. 

A runtime system is automatically linked to the overlay by the compiler. The runtime system is composed of command queues and a task graph and scheduler. Each IP has a command queue containing tasks to be executed. The dependences between any two tasks are represented by a task graph and managed by a scheduler dynamically. How to write such a runtime system is a known technique.

\subsection{Example 2: VGG convolutional neural network}
\label{sec:vgg}

\begin{figure}
    \centering
    \includegraphics[width=0.5\textwidth]{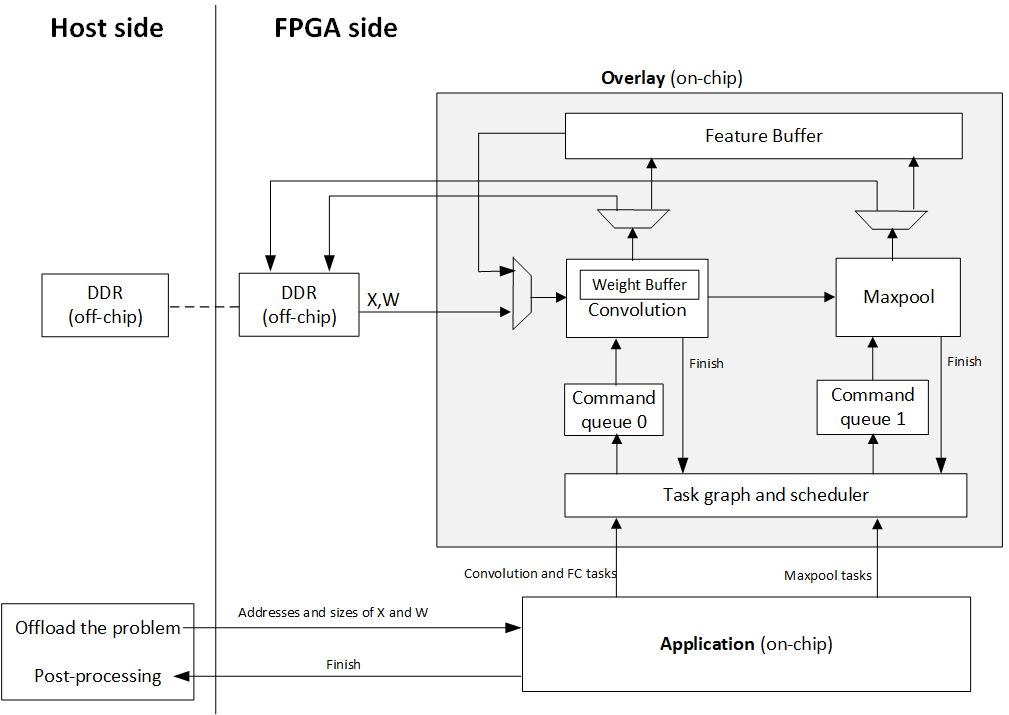}
    \caption{A design for VGG network}
    \label{fig:T2S_Overlay-VGG16-arch}
\end{figure}

A design for VGG is shown in Fig~\ref{fig:T2S_Overlay-VGG16-arch}. There is an overlay and an application on an FPGA. The overlay has 2 hardware IPs: Convolution and Maxpool. All convolution layers (with and without ReLU) and fully-connected (FC) layers can be computed by the Convolution IP, and all the max pooling layers can be computed by the Maxpool IP. The feature map between two layers can be communicated by external DDR, or by an on-chip feature buffer. Inside a layer, the Convolution IP has a weight buffer. 

Algorithm~\ref{alg:vgg} shows for VGG two specifications, following the same principle for the previous LU example. We leave a detailed explanation to the comments there.

\section{Conclusion and Future Work}

We have proposed an idea for a software programmer to quickly build an application-specific overlay on an FPGA, using high-level customizable IPs. We have illustrated the idea with LU decomposition and VGG convolutional neural network. We are building a system to implement the proposed idea, leveraging our previous work on T2S. We will report the progress in future publications.

\begin{algorithm*}[!htb]
\begin{lstlisting}[
    language=C,basicstyle=\tt,numbers=left,stepnumber=1,showstringspaces=false,tabsize=1,breaklines=true,escapechar=`]
    /* Specification 1: Define an overlay */
    Func Convolution(Place::Device), Maxpool(Place::Device);   // Two HW IPs on the device (FPGA)  
    ImageParam X(Float(32), 3), Y(Float(32), 3);               // Input and output feature map
    ImageParam W(Float(32), 4);                                // Weights
    Expr       read_input_from_buffer, store_output_to_buffer, // Control signals to reconfigure
               with_ReLU, is_FC_layer;                         // the overlay. 

    Functional notations and spatial mapping for the Funcs     // Expressible in known state-of-
                                                               // art`~\cite{T2SFCCM19,Susy}`. Details skipped.
        
    // Define a programming interface for each HW IP. Each IP is driven by a command queue.
    // The command queues are automatically provided by the runtime.
    Convolution.command(0, X, Y, W, read_input_from_buffer, store_output_to_buffer, with_ReLU,
                        is_FC_layer);
    Maxpool.command(1, Y, store_output_to_buffer)
    
    // Put the IPs into an overlay, and compile to a named bitstream.
    Overlay(Convolution, Maxpool).compile("overlay.aocx");
\end{lstlisting}
\begin{lstlisting}[
    language=C,basicstyle=\tt,numbers=left,stepnumber=1,showstringspaces=false,tabsize=1,breaklines=true,escapechar=`]
    /* Specification 2: Define an application on the overlay */
    ImageParam X(Float(32), 4), // Input feature map
               Y(Float(32), 4), // Output feature map after the last FC layer.
               W01(Float(32), 5), W23(Float(32), 5),   // Weights for convolution layer 0-1, 2-3
               W46(Float(32), 5), W79(Float(32), 5),   // Weights for convolution layer 4-6, 7-9
               W1012(Float(32), 5),                    // Weights for convolution layer 10-12
               WFC0(Float(32), 2), WFC1(Float(32), 2), // Weights for FC layer 0 and 1
               WFC2(Float(32), 2);                     // Weights for FC layer 2
    Func ConvLayers[13](Place::Device),                // 13 convolution layers
         FCLayers[3](Place::Device),                   // 3 fully connected layers
         MaxpoolLayers[5](Place::Device);              // 5 max pooling layers

    Overlay overlay = load_overlay("overlay.aocx"); // Offload the overlay bitstream to FPGA, if
                                                    //  not yet, and return a handle
         
    #define INPUT(f, i)  f.cropped(3, i, 1)//Reference to i'th channel of the input feature map f
    #define WEIGHT(w, i) w.cropped(4, i, 1)//Reference to the i'th set of weights
    #define DUMMY_I      X                 //Arbitrary input. Not to be used anyway.
    #define DUMMY_O      Y                 //Arbitrary output. Not to be used anyway.
    #define YES          true        
    #define NO           false        

    // Push to queue 0 a convolution task that reads input from DDR, stores output to the feature  
    // buffer, with ReLU, and not a FC layer.
    ConvLayers[0](i)=overlay.enqueue(0, INPUT(X, i), DUMMY_O, WEIGHT(W01, 0), NO, YES, YES, NO);

    // Next layer. Similar to the first layer, but reads from the feature buffer
    ConvLayers[1](i)=overlay.enqueue(0, DUMMY_I, DUMMY_O, WEIGHT(W01, 1), YES, YES, YES, NO); 
    
    // Push to queue 1 a Maxpool task. A Maxpool task always reads input from the feature buffer.
    // Here the task stores output to the feature buffer as well. 
    MaxpoolLayer[0](i)=overlay.enqueue(1, DUMMY_O, YES);

    ConvLayers[2](i)=overlay.enqueue(0, DUMMY_I, DUMMY_O, WEIGHT(W23, 0), YES, YES, YES, NO);
    ConvLayers[3](i)=overlay.enqueue(0, DUMMY_I, DUMMY_O, WEIGHT(W23, 1), YES, YES, YES, NO);
    MaxpoolLayer[1](i)=overlay.enqueue(1, DUMMY_O, YES);
    ConvLayers[4](i)=overlay.enqueue(0, DUMMY_I, DUMMY_O, WEIGHT(W46, 0), YES, YES, YES, NO);
    ConvLayers[5](i)=overlay.enqueue(0, DUMMY_I, DUMMY_O, WEIGHT(W46, 1), YES, YES, YES, NO);
    ConvLayers[6](i)=overlay.enqueue(0, DUMMY_I, DUMMY_O, WEIGHT(W46, 2), YES, YES, YES, NO);
    MaxpoolLayer[2](i)=overlay.enqueue(1, DUMMY_O, YES);
\end{lstlisting}
\caption{Example specifications for VGG.}
\label{alg:vgg}
\end{algorithm*}

\begin{algorithm*}[!htb]
\begin{lstlisting}[
    language=C,basicstyle=\tt,numbers=left,stepnumber=1,firstnumber=41,showstringspaces=false,tabsize=1,breaklines=true,escapechar=`]
    ConvLayers[7](i)=overlay.enqueue(0, DUMMY_I, DUMMY_O, WEIGHT(W79, 0), YES, YES, YES, NO);
    ConvLayers[8](i)=overlay.enqueue(0, DUMMY_I, DUMMY_O, WEIGHT(W79, 1), YES, YES, YES, NO);
    ConvLayers[9](i)=overlay.enqueue(0, DUMMY_I, DUMMY_O, WEIGHT(W79, 2), YES, YES, YES, NO);
    MaxpoolLayer[3](i)=overlay.enqueue(1, DUMMY_O, YES);
    ConvLayers[10](i)=overlay.enqueue(0, DUMMY_I, DUMMY_O, WEIGHT(W1012, 0), YES, YES, YES, NO);
    ConvLayers[11](i)=overlay.enqueue(0, DUMMY_I, DUMMY_O, WEIGHT(W1012, 1), YES, YES, YES, NO);
    ConvLayers[12](i)=overlay.enqueue(0, DUMMY_I, DUMMY_O, WEIGHT(W1012, 2), YES, YES, YES, NO);
    MaxpoolLayer[4](i)=overlay.enqueue(1, Y, NO);//The last MaxPool layer store results to DDR
    
     // An FC layer always reads input from DDR, and stores output to DDR.
    FCLayers[0](i)=overlay.enqueue(0, Y, Y, WFC0, NO, NO, YES, YES);
    FCLayers[1](i)=overlay.enqueue(0, Y, Y, WFC1, NO, NO, YES, YES);
    FCLayers[2](i)=overlay.enqueue(0, Y, Y, WFC2, NO, NO, YES, YES);
    
    // Specify dependences between tasks in different command queues. 
    // Tasks in the same queue are executed in order.
    MaxpoolLayer[0].depend(ConvLayers[1], 0);//Maxpool layer 0 depends on convolution layer 1 
                                             //with distance=0(i.e. in the same loop iteration).
    ConvLayers[2].depend(MaxpoolLayer[0], 0);
    MaxpoolLayer[1].depend(ConvLayers[3], 0);
    ConvLayers[4].depend(MaxpoolLayer[1], 0);
    MaxpoolLayer[2].depend(ConvLayers[6], 0);
    ConvLayers[7].depend(MaxpoolLayer[2], 0);
    MaxpoolLayer[3].depend(ConvLayers[9], 0);
    ConvLayers[10].depend(MaxpoolLayer[3], 0);
    MaxpoolLayer[4].depend(ConvLayers[12], 0);
    FCLayers[0].depend(MaxpoolLayer[4], 0);

    // Set input, compile and run
    set X, W*, and WFC* with real data, and allocate Y a space.
    Target target = get_host_target();      // Get the CPU
    target.set_feature(Target::IntelFPGA);  // The CPU has a FPGA device
    FCLayers[2].realize(n, target);         // Compile all the Funcs into a bitstream, offload
                                            // and run on the FPGA. Here n is #input feature maps
                                            
    // Y contains the results of the final FC layer. The results can be post-processed on the 
    // host side for softmax.
\end{lstlisting}
\end{algorithm*}

\clearpage
\bibliographystyle{abbrv}
\bibliography{main}

\end{document}